# Sodium Pentazolate: a Nitrogen Rich High Energy Density Material

Brad A. Steele, Ivan I. Oleynik*

*Department of Physics, University of South Florida, Tampa, FL 33620*

**Abstract**

Sodium pentazolates $NaN_5$ and $Na_2N_5$, new high energy density materials, are discovered during first principles crystal structure search for the compounds of varying amounts of elemental sodium and nitrogen. The pentazole anion ($N_5^-$) is stabilized in the condensed phase by sodium $Na^+$ cations at pressures exceeding 20 GPa, and becomes metastable upon release of pressure. The sodium azide ($NaN_3$) precursor is predicted to undergo a chemical transformation above 50 GPa into sodium pentazolates $NaN_5$ and $Na_2N_5$. The calculated Raman spectrum of $NaN_5$ is in agreement with the experimental Raman spectrum of a previously unidentified substance appearing upon compression and heating of $NaN_3$.

## Introduction

Nitrogen-rich high energy density materials (HEDMs) [1–4] are being explored as new propellants and explosives. Their high energy content is due to the large amount of energy released upon decomposition of single- and/or double-bonded nitrogen in the condensed phase into triple-bonded gas-phase diatomic $N_2$ molecules. Solid phases of nitrogen at low pressure and temperature are composed of weakly interacting $N_2$ molecules, leading to a plethora of stable molecular crystal phases [5, 6]. It was predicted in 1992 that at 50 GPa the $\epsilon$-$N_2$ molecular crystal transforms into the single-bonded cubic-gauche (cg) polymeric phase of nitrogen [7], which was eventually synthesized by Eremets *et al* [8]. Most recently, a layered polymeric nitrogen crystal has also been discovered [9]. However, the synthesis of these materials occurs at very high pressures ($\sim$ 100 GPa) and temperatures ($\sim$ 2,000 K) and the recovery at ambient conditions was found to be problematic [8–10]. To be useful in real-world applications, HEDMs must at least be metastable at ambient conditions. Therefore new high-nitrogen content materials are actively being searched [1–4].

One promising idea is to achieve metastability of nitrogen-rich HEDMs by introducing foreign atomic impurities into the nitrogen system to facilitate their synthesis, as well as to enhance stability at ambient conditions. For example, the addition of metallic elements to the pure nitrogen system causes electron transfer from metallic to nitrogen atoms. The competition between ionic and covalent bonding might promote new nitrogen phases other than triply bonded $N_2$. The appreciable degree of ionicity might also help to stabilize the new nitrogen condensed phases at ambient conditions.

The high-pressure synthesis of novel high-nitrogen HEDMs can be facilitated by compressing group-I metallic azide precursors, such as sodium azide ($NaN_3$) [11, 12]. An azide is a double bonded linear anion consisting of three nitrogen atoms. The presence of double bonds is believed to assist in either the conversion of azides into extended single-bonded nitrogen structures, or the formation of multi-atom nitrogen clusters. An indirect sign of such processes was seen previously in Raman measurements performed on sodium azide compressed to above 50 GPa [11]. The new peaks in the 700-800 cm$^{-1}$ and 1,000-1,300 cm$^{-1}$ range appearing upon compression and laser heating, cannot be attributed to molecular vibrations of the initial azide precursor [13]. The authors assumed that these new peaks originate from either polymeric nitrogen or a compound containing nitrogen molecular clusters [11]. However the exact crystal structure was never determined.

Recent advances in structure prediction methods

---

*Corresponding author
Email address: oleynik@usf.edu (Ivan I. Oleynik)

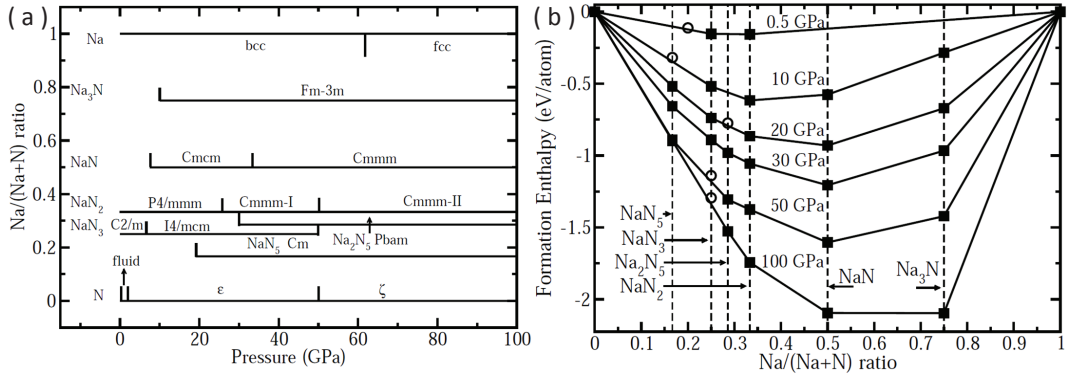

Figure 1: (a) Pressure-composition phase diagram of new Na-N crystal phases discovered in simulations. (b) Convex hull diagram at pressures 0.5, 10, 20, 30, 50 and 100 GPa. Solid squares represent thermodynamically stable phases, open circles – metastable phases.

[14–17] make it possible to screen compounds of all possible stoichiometries and find the most stable structures using first-principles density functional theory. Being motivated by the goal of discovery of high-nitrogen content HEDM's stabilized by metallic cations, we undertook a systematic search of $Na_xN_y$ materials containing variable number of elemental sodium and nitrogen atoms in the crystal lattice. The calculated enthalpies of the predicted structures are used to determine the most stable compounds at a given pressure, allowing us to construct the phase diagram of the materials of varying stoichiometry. Although the structure search has not been specifically biased towards systems containing $N_5^-$, new compounds, sodium pentazolates $NaN_5$ and $Na_2N_5$, metastable at ambient conditions, are discovered. The detailed characterization of these materials is performed by comparing the calculated Raman spectra with those obtained in experiment [11], to answer the important question whether the sodium pentazolates appear upon compression and heating of sodium azide precursor.

### Computational details

The search for new $Na_xN_y$ compounds of varying stoichiometry is performed at varying pressures by using the first principles evolutionary structure prediction method USPEX [14–16]. The unit cell for the structure search contains between 6-16 atoms which covers a substantially large portion of the energy landscape. The prediction of new structures at a given pressure begins by generating crystal structures with randomized chemical composition, lattice parameters, and atomic coordinates, followed by energy minimization using density functional theory (DFT). The structures are ranked according to their enthalpies of formation, then variation operators are applied to the best of them having lowest enthalpies to generate structures of a new generation. The process repeats until the lowest enthalpy crystals do not change for several generations, thus indicating that no new structures with lower enthalpies will appear. For a given pressure, the convex hull, a formation enthalpy-composition curve is constructed. It consists of points that correspond to the structure with the lowest formation enthalpy at a given composition with respect to the lowest enthalpy structures of the base elements. The reference structures are $\alpha$-$N_2$, $\epsilon$-$N_2$, and cg-N for nitrogen and bcc-Na and fcc-Na for sodium each taken at corresponding pressure.

First-principles calculations are performed using the Perdew-Burke-Ernzerhof (PBE) generalized gradient approximation (GGA) functional [18] within density functional theory (DFT) implemented in VASP [19] and DMol [20]. The PBE functional has been previously shown to give reliable results for sodium azide[13, 21]. In VASP calculations, ultrasoft pseudopotentials [22] with inner core radii of 1.757 Å for Na and 0.873 Å for N and plane wave basis sets are used with an energy cutoff of 600 eV and a 0.05 Å$^{-1}$ k-point sampling. Charges on atoms and bond orders are calculated with DMol, the latter is also used to perform molecular dynamics simulations at high temperatures to establish dynamical stability of new compounds. Vibrational properties including phonon dispersion curves are calculated– using the PHONOPY code [23], which interfaces with VASP.

In addition, the adequate accuracy in predicting the small enthalpy differences is established by comparing the formation enthalpies calculated by the PBE GGA and HSE06 hybrid functionals. The formation enthalpy for each crystal include zero-point energy contributions that are calculated using the vibrational spectra. The calculations using the HSE06 functional use projector augmented wave (PAW) pseudopotentials [24] with inner core radii of 1.757 Å for Na and 0.582 Å for N. A large energy cutoff of 1,000 eV is used to accurately describe the wavefunction near the nitrogen core. Convergence studies show that only a 0.10 Å$^{-1}$ k-point sampling is adequate. Dmol all-electron calculations of charges and bond order are performed using DND local combination of atomic orbitals (LCAO) basis sets.

### Results and Discussion

The robustness of the algorithm for discovering new crystal phases of materials with varying stoichiometries is demonstrated by predicting known phases of sodium azide (NaN$_3$) without any prior input. The structure search correctly finds the α-phase of NaN$_3$ with the symmetry C2/m at 0.5 GPa in agreement with experiment, while at 30 GPa it correctly produces the I4/mcm polymorph of NaN$_3$ [21, 25]. At 60 GPa, the P6/m-Na$_2$N$_6$ structure featuring N$_6$ rings is found to be the lowest enthalpy phase with a 1:3 sodium to nitrogen ratio, consistent with previous calculations [21].

In addition to known phases of NaN$_3$, several new crystals with novel stoichiometries are discovered, their crystal structure information being provided in Supplementary Table S1. In Figure 1(a) and (b), the predicted phase diagram, as well as the calculated convex hull, at several pressures are shown. The crystal structures of materials with variable composition are labeled by the symmetry followed by the chemical formula. Figure 2 displays the crystal structures of several newly discovered materials.

In order to further justify the adequate performance of PBE functional in predicting accurate formation enthalpies, the convex hull at 50 GPa has been calculated using both PBE functional and the hybrid HSE06 functional [26], see comparison in

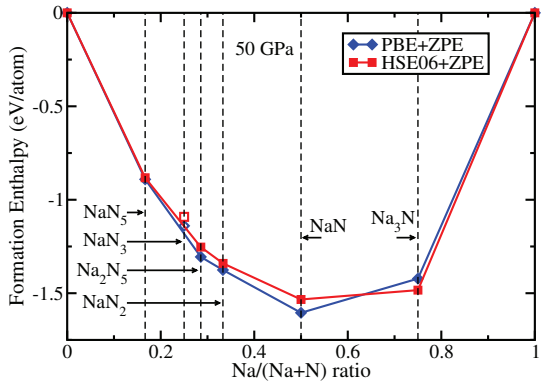

Figure 3: A comparison between the convex hull's calculated with PBE and HSE06 at 50 GPa with zero-point energies (ZPE) included.

Figure 3. The HSE06 formation enthalpies for most of the compounds are slightly higher than those calculated using PBE functional, with exception of Na$_3$N which has lower HSE06 formation enthalpy. Overall, both HSE06 and PBE convex hulls are very similar. The HSE06 functional is considered to be state-of-the-art and gives formation enthalpies and atomization energies close to experiment across a wide range of molecules and crystals [27]. Therefore, the similarity of the two curves demonstrates a good accuracy of the PBE calculations of the systems under study.

It is found that NaN$_3$ no longer resides on the convex hull above 50 GPa, thus implying that it is thermodynamically unstable beyond this pressure. This results is also reproduced with the HSE06 functional, see Figure 3. Therefore, upon compression in a diamond anvil cell (DAC) above 50 GPa, NaN$_3$ will transform into a combination of Na$_x$N$_y$ phases.

The crystal phases P2/c-NaN$_5$, Cm-NaN$_5$, and Pbam-Na$_2$N$_5$, shown in Figure 2(a-c), are the most interesting of all the newly discovered materials since they contain pentazole molecules that can be potential sources of unidentified Raman peaks in experiments on compression of NaN$_3$ at high pressures [11]. Our calculations predict that Cm-NaN$_5$ and Pbam-Na$_2$N$_5$ are thermodynamically stable above 20 GPa and 30 GPa respectively; see Figure 1 (b). These two structures may be synthesized by compressing NaN$_3$ above 50 GPa, since NaN$_3$ becomes unstable above this pressure. Alternatively, direct synthesis of NaN$_5$ can be facilitated by heating and compressing $NaN_3$ in a DAC using a nitrogen-rich pressure medium via the following chemical reaction, $N_2+NaN_3\rightarrow NaN_5$.

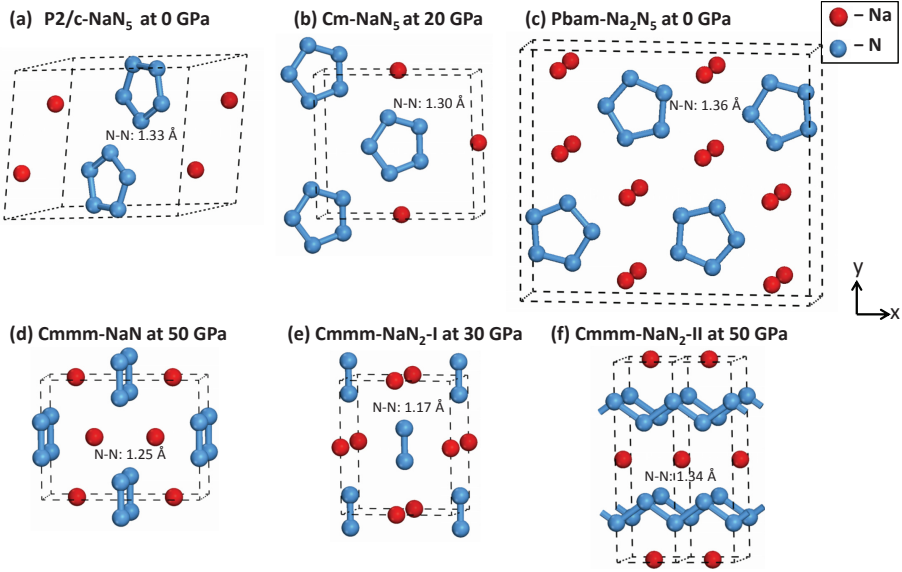

Figure 2: Several newly-predicted $Na_xN_y$ crystals at pressures corresponding to where they are thermodynamically stable, or in the case of $P2/c$-$NaN_5$ and $Pbam$-$Na_2N_5$ metastable at 0 GPa. The space group-composition of each structure is shown as well as the N bond length.

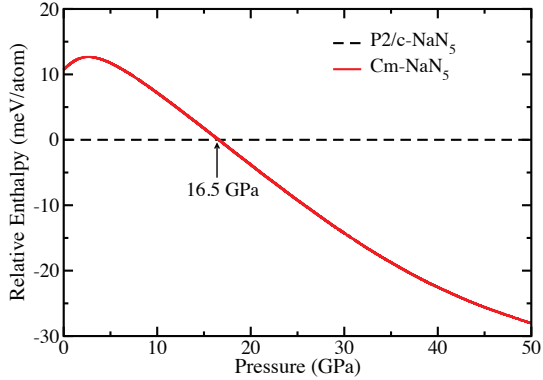

Figure 4: The relative enthalpy of the two phases with $NaN_5$ stoichiometry.

In order for these materials to be useful as HEDMs they must at least be metastable at ambient conditions. To test for the metastability of new materials, the most thermodynamically stable polymorphs of $NaN_5$ and $Na_2N_5$ at ambient conditions are determined using the USPEX method. At 0 GPa a new polymorph, $P2/c$-$NaN_5$ is found to be of the lowest enthalpy compared to $Cm$-$NaN_5$, see Figure 2(a), while $Pbam$-$Na_2N_5$, see Figure 2(c), remains the only polymorph of $Na_2N_5$, metastable at ambient conditions as well as at higher pressures. The transition pressure between the high-pressure $Cm$ phase of $NaN_5$ ($Cm$-$NaN_5$) and the zero pressure phase ($P2/c$-$NaN_5$) is calculated to be 16.5 GPa, see Figure 4.

The dynamical stability of the crystal phases $P2/c$-$NaN_5$ and $Pbam$-$Na_2N_5$ is determined using two independent methods. First, the phonon dispersion curves are calculated and show the absence of any imaginary frequencies at 0 GPa in the entire Brillouin zone; see Figure 5(a,b). As an independent check of the dynamical stability of these two phases at 0 GPa, DFT molecular dynamics simulations at 1,000 K are also performed in the NVT ensemble with a 0.5 fs time step for 10 ps and no chemical decomposition is observed. This implies that both of these phases are dynamically stable at 0 GPa.

The phonon frequencies at the gamma point, shown in Figure 5(a,b), can be useful in identifying these newly predicted phases using vibrational spectroscopic methods such as Raman and infrared spectroscopy. The modes in the interval 80-250 cm$^{-1}$ are the lattice modes as well as pentazole librational modes in both phases. The modes at 635 cm$^{-1}$ in Pbam-Na$_2$N$_5$ and 766 cm$^{-1}$ in P2/c-NaN$_5$ are bending modes of the pentazole molecule. The modes near 900 cm$^{-1}$ in Pbam-Na$_2$N$_5$ and 990 cm$^{-1}$ in P2/c-NaN$_5$ are deformational modes of the pentazole molecule. At higher frequencies are the breathing modes of the pentazole molecule with the symmetric stretch located at 1,050 cm$^{-1}$ in Pbam-Na$_2$N$_5$ and 1,170 cm$^{-1}$ in P2/c-NaN$_5$. All the frequencies given here are at 0 GPa.

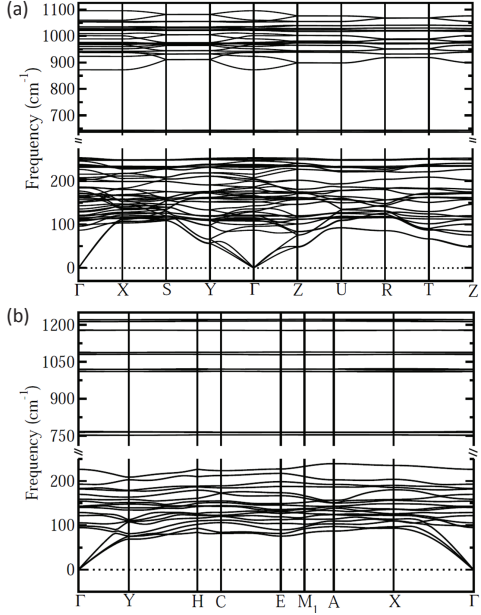

Figure 5: Phonon dispersion curves for the crystal phases (a) Pbam-Na$_2$N$_5$ and (b) P2/c-NaN$_5$ at 0 GPa. The lack of imaginary frequencies in the full Brillouin zone shows both phases that are dynamically stable at 0 GPa.

Table 1: Calculated Mulliken charges and Mayer Bond Orders for the nitrogen atoms and bonds in each new crystal at 0 GPa.

| Structure | N-cluster | Charge | Bond Order |
|---|---|---|---|
| P2/c-NaN$_5$ | N$_5$ | -0.166 | 1.42 |
| Pbam-Na$_2$N$_5$ | N$_5$ | -0.306 | 1.20 |
| P4/mmm-NaN$_2$ | N$_2$ | -0.340 | 2.09 |
| Cmmm-NaN$_2$-II | N-chain | -0.320 | 1.17 |
| Cmcm-NaN | N$_2$ | -0.630 | 1.46 |

The bonding in the pentazole anions in the NaN$_5$ and Na$_2$N$_5$ crystal phases is aromatic, which is reflected in the calculated bond lengths and bond orders. The N–N bond lengths in the pentazoles are between the N–N single bond (1.449 Å as in hydrazine [28]) and double bond ( 1.252 Å as in trans-diimine, [29]); see Figure 2(a-c). The calculated bond orders are also between the single (1.0) and double (2.0) bond; see Table 1. The charge (-0.83 e) and the bond order (1.42) are also close to those found in the gas-phase N$_5^-$ anion (-1 e and 1.45 respectively), which shows the structural and chemical similarity of the N$_5^-$ anion in both gas phase and crystalline NaN$_5$ environments. The calculated band structure demonstrates that P2/c-NaN$_5$ is an insulator with a band gap of 5 eV while Pbam-Na$_2$N$_5$ is metallic (see Supplementary Figures. S6 and S7).

In addition to the discovery of a new class of pentazolate compounds, several other new crystal phases with unusual physical properties are found during the structure search. These crystal structures contain either diatomic N$_2$ anions or ionic nitrogen chains, as well as Na cations, see Figure 2(d-f) for a few representative structures and Figure 1(a,b) for their thermodynamic stability range. The crystals containing N$_2$ anions are found in both the NaN$_2$ and NaN stoichiometries with differing degrees of ionicity and bond orders. The calculated charges on each N atom and the N-N bond order are provided in Table 1. In the case of P4/mmm-NaN$_2$ and Cmmm-NaN$_2$, the N$_2$ anion has a negative charge slightly less than −1.0 e while the N-N bond order 2.09 is close to that of a double bond, which is in contrast to zero charge and bond order 3.0 in the case of the gas phase N$_2$ molecule. The NaN materials (Cmcm-NaN and Cmmm-NaN) have a larger negative charge on the N$_2$ anion as compared to the P4/mmm-NaN$_2$ and Cmmm-NaN$_2$ cases, which lowers the bond order to 1.46 due to electronic filling of the anti-bonding states. At pressures exceeding 50 GPa the lowest enthalpy NaN$_2$ structure consists of infinite zig-zag nitrogen chains in the space group Cmmm, referred to as Cmmm-NaN$_2$-II, as shown in Figure 2(f). The calculated band structure for this material indicates it is also metallic with occupied states at the Fermi level as shown in Supplementary Figure S4. The metallic nature of this material is consistent with a similar metallic structure, predicted previously, which also consists of infinite zig-zag nitrogen chains in the Cmcm space group [30]. The Cmmm-NaN$_2$-II structure thermally decomposes in DFT molecular dynamics simulations performed at 0 GPa and 1,000 K. This means that this crystal is dynamically unstable at 0 GPa and therefore is not a suitable candidate for experimental synthesis.

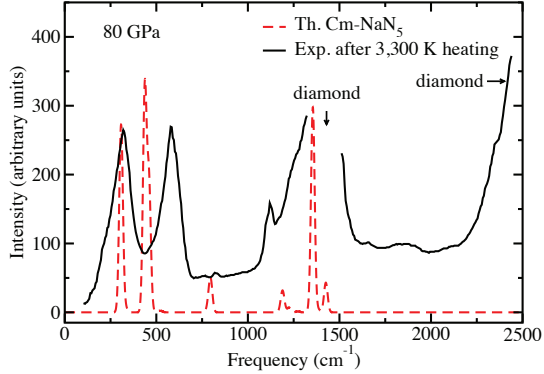

Figure 6: Theoretical Raman spectrum of Cm-NaN$_5$ (red dashed line) compared to the experimental Raman spectrum [11](black solid line). The calculated N$_2^-$ stretching frequency for Cmmm-NaN$_2$-I is marked by an arrow.

There is substantial evidence for the appearance of the new sodium pentazolate structures predicted in this work in experiments by Eremets et al [11]. They compressed and heated sodium azide to high pressures and temperatures. At pressures above 80 GPa, and a temperature of about 3,000 K, the peaks in the Raman spectrum associated with sodium azide were found to disappear. This is in agreement with the theoretical prediction made in this paper that sodium azide becomes thermodynamically unstable at high pressures. More significantly, the Raman spectrum of the newly predicted sodium pentazolate phase Cm-NaN$_5$ shows good agreement with that in experiment; see Figure 6. In particular, the agreement is best near 760 cm$^{-1}$ (the pentazole bending mode) and near 1,150 cm$^{-1}$ (the pentazole deformation mode). Unfortunately, interference from the diamond first order peak inhibits the detection of the pentazole stretching mode, see Figure 6. Also, in agreement with experiment, the Cm-NaN$_5$ phase has two lattice modes with appreciable intensity at 307 cm$^{-1}$ and 440 cm$^{-1}$ compared to the experimental frequencies at 320 cm$^{-1}$ and 580 cm$^{-1}$. The relatively large difference between the experimental and theoretical frequencies of the second peak may be due the non-hydrostatic effects at 80 GPa which are not considered in our calculations. In addition, it is highly unlikely that NaN$_3$ will be fully converted into the pure NaN$_5$ compound. Therefore a one-to-one correspondence between theory and experiment is not expected due to possible appearance of unidentified nitrogen-containing species as well as bulk sodium upon compression and heating of NaN$_3$.

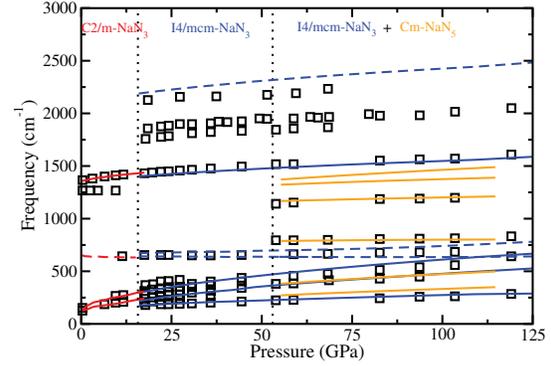

Figure 7: Evolution of theoretical NaN$_5$ and NaN$_3$ Raman active (solid lines) and IR active (dashed lines) modes with pressure compared to experiment [11] (black open squares). Agreement between theoretical and experimental frequencies above 50 GPa suggests the synthesis of the NaN$_5$ compound.

The agreement between the theoretical and experimental frequencies of the Raman-active modes of NaN$_5$ at room temperature as a function of pressure are also remarkably good, see Figure 7. As shown in Figure 7, the internal Raman active modes of the pentazole in NaN$_5$ with frequencies 760 cm$^{-1}$ and 1,150 cm$^{-1}$ appear in the experimental spectra above 50 GPa that are not from the initial azide compound. This is in agreement with the prediction that NaN$_5$ becomes stable above 20 GPa. Since these measurements are performed by compressing NaN$_3$ at room temperature without additional heating, it expected that complete conversion of NaN$_3$ into NaN$_5$ is not achieved. That is why the peaks from sodium azide (NaN$_3$) are still present in the experimental spectrum up to 100 GPa, see Figure 7. Overall, the agreement between theory and experiment suggests the synthesis of the predicted NaN$_5$ compound.

The major nitrogen unit in sodium pentazolates NaN$_5$ and Na$_2$N$_5$ is the cyclic pentazole anion N$_5^-$. There has been previous work on the synthesis of the gas phase $N_5^-$ by cleavage of the C-N bond in substituted phenylpentazoles [31], and using laser desorption ionization of dimethyl-aminophenylpentazole (DMAPP) [32]. Theoretical calculations also support the formation of the gas phase $N_5^-$ from substituted phenylpentazoles [33]. The gas phase $Na^+N_5^-$ has also been studied theoretically, and it was found that it has an activation barrier large enough to make it possible to exist at ambient pressure [34]. However, previous calculations have yet to explore $NaN_5$ crystals at high pressure, such as what is studied in this work.

Since bonding in the pentazoles is aromatic with a bond order between $1.0 - 2.0$, the energy release upon completion of detonation chemistry is expected to be substantial. The detonation performance of an HEDM is determined by several factors, including a large positive enthalpy of formation. The formation enthalpies at ambient pressure, 84.40 kJ/mol for P2/c-$NaN_5$ and 131.57 kJ/mol for Pbam-$Na_2N_5$, calculated using the HSE06 hybrid functional, are comparable and even higher than those of the commonly used HEDMs, such as RDX and HMX: 61.55 kJ/mol and 75.02 kJ/mol respectively [35]. Although a detailed investigation of detonation thermochemistry is beyond the scope of this work, the large formation enthalpy indicates that $NaN_5$ and $Na_2N_5$ are powerful HEDMs.

## Conclusions

In conclusion, first-principles evolutionary structure search of compounds consisting of sodium and nitrogen uncovered several new high-nitrogen content materials. Two of them, Cm-$NaN_5$ and Pbam-$Na_2N_5$, featuring pentazole anions in the crystalline state, are thermodynamically stable at high pressures, their zero pressure polymorphs, P2/c-$NaN_5$ and Pbam-$Na_2N_5$, being metastable at ambient conditions. The sodium pentazolates $NaN_5$ and $Na_2N_5$ have bond orders between 1.2-1.5, and are made up of 56-76 % nitrogen by weight at ambient conditions. These properties make these newly predicted crystals excellent candidates for HEDMs as the single-bonded pentazoles decompose to triple-bonded gas phase $N_2$ molecules. According to our calculations, the sodium azide precursor becomes metastable above 50 GPa, resulting in its decomposition to the newly predicted crystal structures upon application of pressure. The theoretically predicted Raman spectrum of Cm-$NaN_5$ displays good agreement with experiment. Therefore, these calculations provide evidence that these new compounds were synthesized in previous DAC experiment [11]. Further thorough experimental validation of our theoretical predictions is urgently sought as the potential usefulness and predicted stability of new nitrogen-rich materials warrants such new efforts.

## Acknowledgements

The authors thank Jonathan Crowhurst, Elissaios Stavrou, Joseph Zaug, and Suhithi Peiris for enlightening discussions. This research is supported by Defense Threat Reduction Agency, grant # HDTRA1- 12-1-0023. Calculations were performed using the NSF XSEDE facilities, the USF Research Computing Cluster, and the computational facilities of the Materials Simulation Laboratory at USF.

## Appendix A. Supplementary Data

Supplementary data associated with this article can be found, in the online version, at

# Supporting information for:

# Sodium Pentazolate: a Nitrogen Rich High Energy Density Material

Brad A. Steele and Ivan I. Oleynik*

*Department of Physics, University of South Florida, Tampa, FL 33620*

E-mail: oleynik@usf.edu

# Electronic and Phonon Band Structure of Selected $Na_xN_y$ Crystal Structures

Here the electronic and phonon band structure for several new crystals with $Na_xN_y$ stoichiometries is provided. Between 30-50 GPa, the two phases with the lowest formation enthalpy are the NaN and $NaN_2$ crystals. Interestingly, all consist of diatomic $N_2$ anions and Na cations in this pressure range (30-50 GPa). The phonon dispersion curves for the NaN and $NaN_2$ structures in Fig. S1 and Fig. S2 demonstrate that these phases are dynamically stable at 30 GPa as they lack the modes with imaginary frequencies throughout the entire Brillouin zone. Two NaN phases are found, one in the Cmmm space group as shown in Fig. 2(d) in the main text and another in the Cmcm space group. Neither of the two NaN crystal phases are thermodynamically stable at zero pressure.

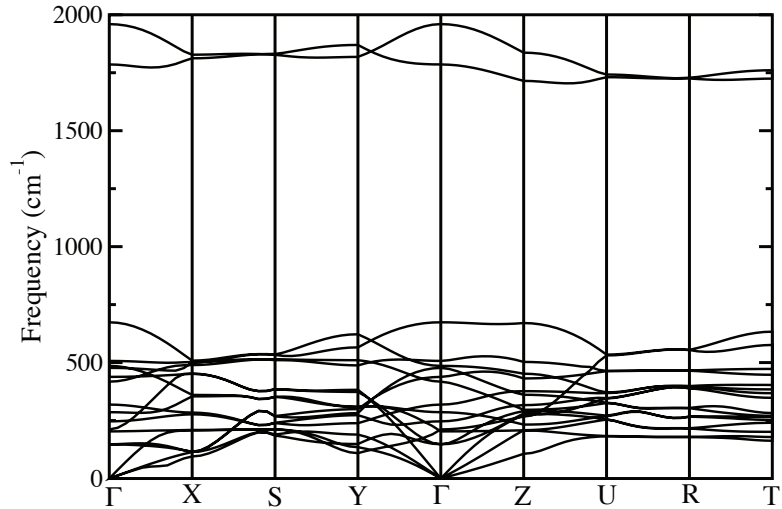

Figure S1: Phonon dispersion curves for Cmmm-$NaN_2$-I at 30 GPa.

It is also found that the frequency of the $N_2$ stretching mode for the $N_2$ anions in P4/mmm-$NaN_2$ and Cmmm-$NaN_2$-I is reduced with respect to the gas phase $N_2$ molecule, see Fig. S1. This type of damped $N_2$ vibration has been suggested in[1] to correspond to the peak near 1,900 cm$^{-1}$ in the experimental Raman spectra of the compression of sodium azides to high pressures inside a DAC[2]. The $N_2$ stretching mode frequency is reduced even

S2

further for the NaN phases; see Fig. S2. However, all of the $N_2$-containing NaN and $NaN_2$ polymorphs are found to be metallic, see the electronic band structure in Figs. S3 and S4. Due to metallicity of these compounds, their appearance in high-pressure synthesis experiments is difficult to assertain as the very small changes in polarizability of the crystal is expected to result in very weak Raman intensities.

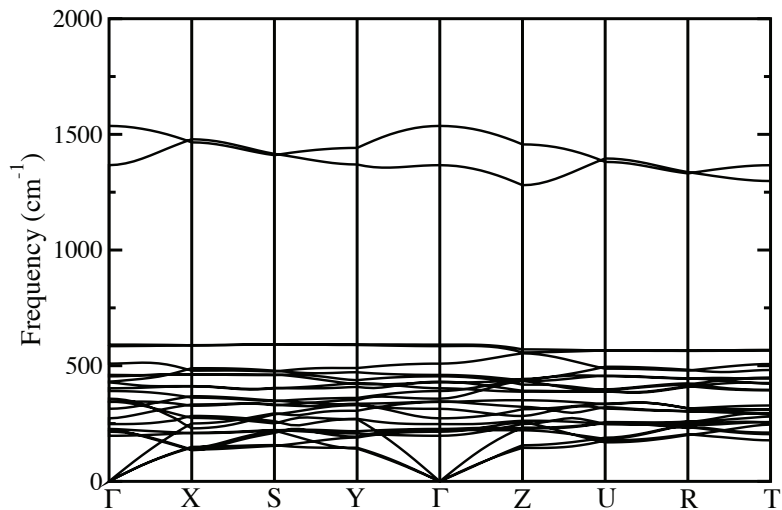

Figure S2: Phonon dispersion curves for Cmmm-NaN at 30 GPa.

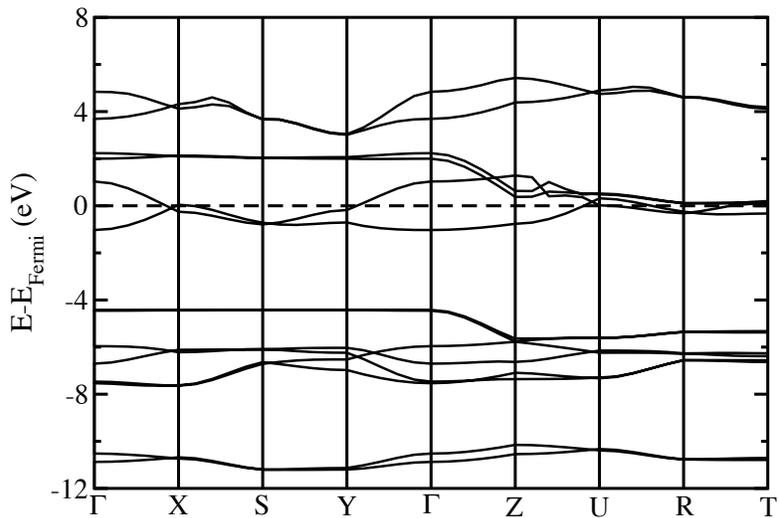

Figure S3: Electronic band structure for Cmmm-NaN at 0 GPa.

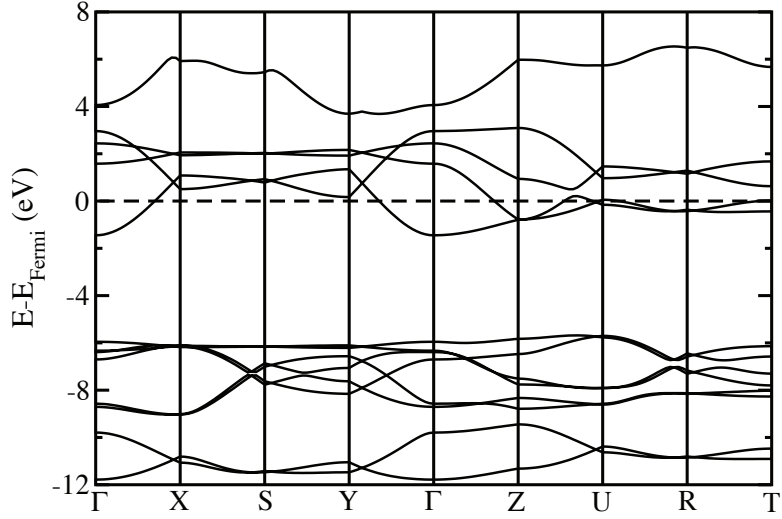

Figure S4: Electronic band structure for Cmmm-NaN$_2$-I at 0 GPa.

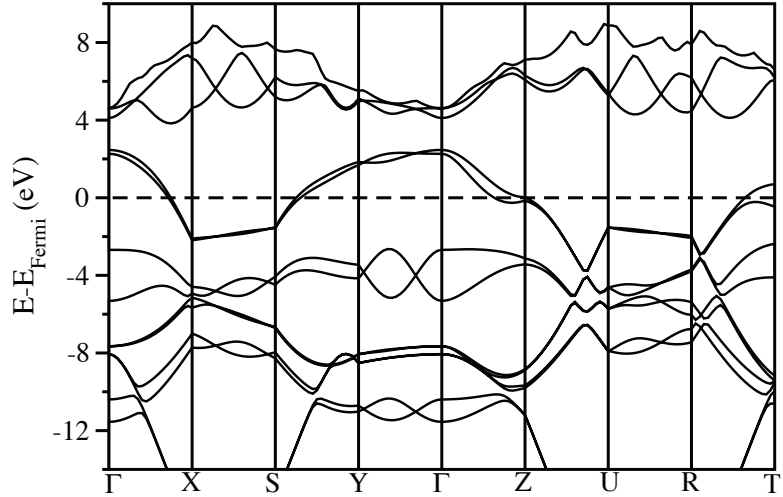

Figure S5: Electronic band structure for Cmmm-NaN$_2$-II at 0 GPa.

The two phases that are considered the most important in this paper are the sodium pentazolates NaN$_5$ and Na$_2$N$_5$. Our electronic structure calculations, shown in Fig. S6, demonstrate that the P2/c-NaN$_5$ phase is an insulator with a band gap of about 5 eV calculated using the PBE functional. The electronic band structure for Cm-NaN$_5$, which is the high-pressure polymorph of P2/c-NaN$_5$, is qualitatively the same with about the same band gap. The Pbam-Na$_2$N$_5$ phase is predicted to be metallic, as shown in Fig. S7. Therefore, the NaN$_5$ phase is the only insulating phase, other than pure nitrogen structures such as $N_2$ molecular crystals or cg-N, that are predicted to be thermodynamically stable above 50 GPa. Hence, it is anticipated that other than pure nitrogen, Cm-$NaN_5$ will make to most substantial contribution to the Raman spectrum at pressures above 50 GPa (the pressure at which sodium azide is predicted to become unstable).

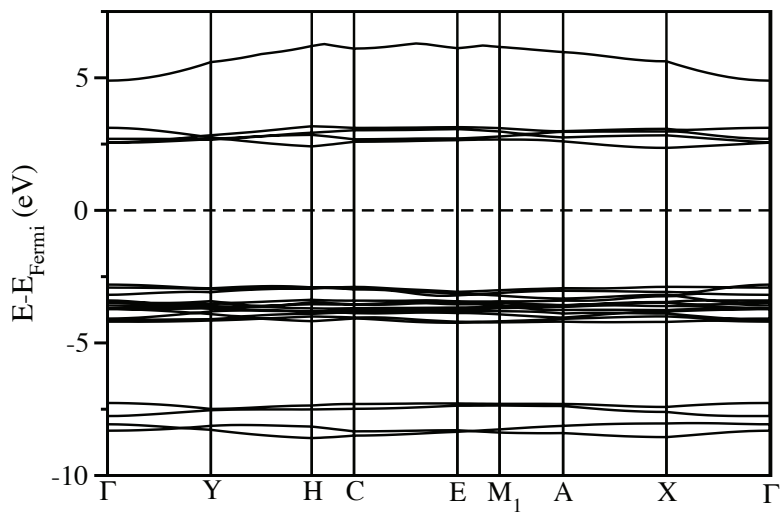

Figure S6: Electronic band structure for P2/c-$NaN_5$ at 0 GPa.

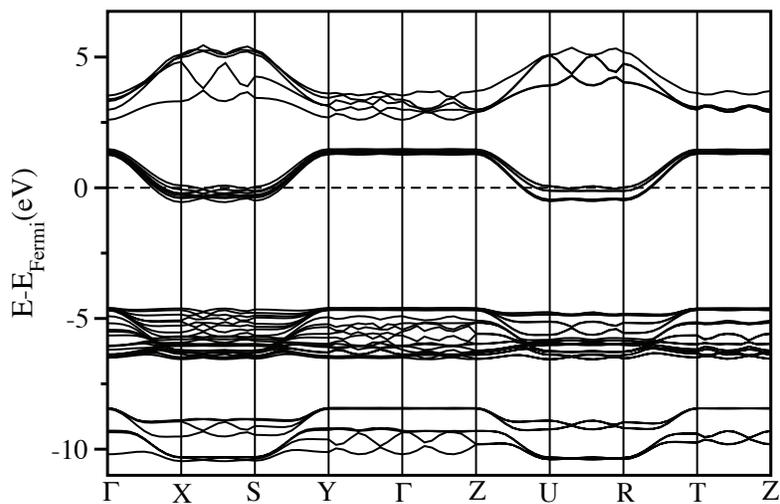

Figure S7: Electronic band structure for Pbam-$Na_2N_5$ at 0 GPa.

# Crystal Structure for Selected $Na_xN_y$ Crystal Phases

The crystal structural information of several new phases with $Na_xN_y$ stoichiometries is provided in Table S1. The pressures at which the structures are given correspond to where they are thermodynamically stable except for P2/c-$NaN_5$ and Pbam-$Na_2N_5$, which are given at 0 GPa because they are dynamically stable there.

Table S1: Crystal structure for several newly predicted structures at the corresponding pressure at which they are stable.

| structure | pressure (GPa) | space group | lattice parameters | | | | atoms | atomic coordinates (fractional) | | |
|---|---|---|---|---|---|---|---|---|---|---|
| | | | a(Å) | b(Å) | c(Å) | $\beta(°)$ | | x | y | z |
| Cm-NaN$_5$ | 30 | Cm | 6.267 | 5.445 | 2.971 | 78.07 | Na1 | 0.487 | 0.000 | 0.245 |
| | | | | | | | N1 | 0.840 | 0.000 | 0.922 |
| | | | | | | | N2 | 0.129 | 0.881 | 0.475 |
| | | | | | | | N3 | 0.452 | 0.693 | 0.756 |
| P2/c-NaN$_5$ | 0 | P2/c | 9.065 | 5.574 | 6.868 | 152.639 | Na1 | 0.000 | 0.287 | 0.750 |
| | | | | | | | N1 | -0.721 | 0.274 | 0.411 |
| | | | | | | | N2 | -0.362 | 1.048 | -0.041 |
| | | | | | | | N3 | -0.500 | 0.413 | -0.250 |
| Pbam-Na$_2$N$_5$ | 0 | Pbam | 10.776 | 9.297 | 3.100 | | Na1 | -0.869 | 0.918 | 0.000 |
| | | | | | | | Na2 | -0.137 | 0.411 | 0.000 |
| | | | | | | | N1 | -0.025 | 0.683 | -0.50 |
| | | | | | | | N2 | -0.775 | 0.261 | -0.50 |
| | | | | | | | N3 | -0.158 | 0.864 | -0.50 |
| | | | | | | | N4 | -0.858 | 0.372 | -0.50 |
| | | | | | | | N5 | -0.965 | 0.170 | -0.50 |
| Cmmm-NaN$_2$-I | 30 | Cmmm | 4.078 | 4.854 | 2.621 | | Na1 | -0.50 | 0.00 | 1.00 |
| | | | | | | | N1 | 0.00 | -0.121 | 1.50 |
| Cmmm-NaN$_2$-II | 60 | Cmmm | 2.170 | 7.270 | 2.575 | | Na1 | 0.00 | 0.50 | 0.50 |
| | | | | | | | N1 | 0.50 | 0.697 | 0.00 |
| Cmmm-NaN | 50 | Cmmm | 5.612 | 4.239 | 2.517 | | Na1 | 0.193 | 0.000 | 0.500 |
| | | | | | | | N1 | 0.000 | 0.352 | 0.000 |
| Cmcm-NaN | 30 | Cmcm | 4.494 | 7.669 | 4.068 | | Na1 | 0.000 | -0.188 | 0.750 |
| | | | | | | | Na2 | 0.500 | -0.096 | 0.750 |
| | | | | | | | N1 | -0.357 | -0.606 | 1.250 |